\documentclass[sigconf, authorversion]{acmart}

\setcopyright{none}              
\renewcommand\footnotetextcopyrightpermission[1]{} 

\AtBeginDocument{%
  }

\usepackage[most]{tcolorbox}
\usepackage{enumitem}
\usepackage{booktabs} 
\usepackage{array}
\usepackage{multirow} 
\usepackage{multicol} 
\usepackage{xcolor}
\usepackage[most]{tcolorbox}
\usepackage[normalem]{ulem}

\usepackage{listings}
\newtcblisting{callout}[2][]{%
  colback=gray!5,
  colframe=gray!60,
  boxrule=0.8pt,
  title=#1,
  listing only,
  left=4pt, right=4pt, top=2pt, bottom=2pt,
  enhanced
}

\usepackage{tikz}
\usetikzlibrary{shapes.geometric, arrows.meta, positioning}
\usepackage{xcolor}
\usepackage{pifont}
\usepackage{svg}

\begin{document}

\title[Guardrails or Roadblocks?]{Guardrails or Roadblocks? Effects of Pedagogical Style and Context Awareness in AI Teaching Assistants for Programming}

\author{Madeleine Eastwood}
\affiliation{%
  \institution{University of Florida}
  \city{Gainesville}
  \state{FL}
  \country{United States}
}
\email{meastwood@ufl.edu} 

\author{Harshith Narne}
\affiliation{%
  \institution{University of Florida}
  \city{Gainesville}
  \state{FL}
  \country{United States}
}
\email{h.narne@ufl.edu}

\author{Joseph Hilby}
\affiliation{%
  \institution{University of Florida}
  \city{Gainesville}
  \state{FL}
  \country{United States}
}
\email{joseph.hilby@gmail.com}

\author{Paul Denny}
\orcid{0000-0002-5150-9806}
\affiliation{
  \institution{University of Auckland}
  \city{Auckland}
  \country{New Zealand}
}
\email{paul@cs.auckland.ac.nz}

\author{Ashish Aggarwal}
\orcid{----------------}
\affiliation{%
  \institution{University of Florida}
  \city{Gainesville}
  \state{FL}
  \country{United States}
}
\email{ashishjuit@ufl.edu}

\author{Amanpreet Kapoor}
\orcid{0000-0003-1340-8315}
\affiliation{%
  \institution{University of Florida}
  \city{Gainesville}
  \state{FL}
  \country{United States}
}
\email{kapooramanpreet@ufl.edu}

\begin{abstract}

AI teaching assistants (AI TAs) backed by large language models (LLMs) and pedagogical guardrails are increasingly being integrated into programming courses, providing students with scalable access to hints, conceptual explanations, and code-level feedback. However, guardrails may also create friction. If students feel that the support provided is overly restrictive or poorly contextualized to their current progress, they may bypass approved tools for general-purpose LLMs. To investigate how AI TA design affects students' learning experiences, we conducted a randomized controlled trial with 132 students in an introductory programming course. Students completed three tasks related to code-writing and debugging and were randomly assigned to one of four AI TAs varied across two dimensions: \textit{pedagogical guidance style} (Socratic vs. Direct instruction) and \textit{context awareness} (no context vs. full context of the problem and student solution). We examined students' perceptions, interaction behaviors, and evidence of post-task comprehension. Students rated the Socratic AI TA with full context least favorably, reporting significantly lower perceived support for task completion. Descriptively, this condition also showed the highest observed interaction stress, the highest rate of external LLM use, and the lowest proportion of post-task explanations demonstrating full comprehension, though these differences were not statistically significant. These findings suggest that guardrailed AI TAs are not automatically better for learning. Instead, their effectiveness depends on how pedagogical guidance and contextual awareness are balanced in ways that students experience as useful, supportive, and worth continuing to use.

\end{abstract}



\maketitle

\section{Introduction}

LLM-backed AI teaching assistants (AI TAs) are increasingly being deployed in programming courses to provide scalable, always-available help while using pedagogical guardrails to avoid simply giving students solutions \cite{liffiton_2024_codehelp, kazemitabaar_2024_codeaid, sheese_2024_llm_ta_patterns, 2025GuardrailsBehaviorsKapoor}. Unlike general-purpose LLM chat tools, these systems are designed to support learning by providing tutor-like guidance without directly completing the work for students. Yet these guardrails create a design tension, because if a system is too restrictive or if students perceive the help as poorly contextualized to their request, they may disengage, avoid seeking help, or bypass approved tools in favor of unrestricted general-purpose LLMs \cite{ko_2025_relationship_help_seeking, 2025GuardrailsBehaviorsKapoor}. This creates a problem for educators because when students migrate to external systems, opportunities to capture learning-related data are lost, and students may become more susceptible to over-reliance on AI-generated solutions \cite{bo_2026_saboteurs, denny_2024_computing_education, shein2024impact}. At the same time, student favorability is not necessarily equivalent to learning effectiveness: a guardrailed AI TA that provides more friction between the student and the answer may be perceived as less helpful or favorable to the student, while still supporting the productive struggle associated with deeper learning. The goal, then, is not to maximize favorability alone, but to identify configurations that are both pedagogically sound and experienced by students as useful and worth continuing to use.

A central challenge for our community is to understand how design choices shape students' experiences with AI TAs. Prior work has introduced and evaluated guardrailed AI systems for programming support \cite{liffiton_2024_codehelp, kazemitabaar_2024_codeaid, 2025GuardrailsBehaviorsKapoor}, and instructors have expressed a strong preference for customizing AI chatbots to fit their course-specific needs \cite{hou_2026_bespoke_bots}. However, most prior work has evaluated fixed AI TA systems as complete interventions, without explicitly testing how \emph{different configurations of the same system} affect student perceptions, behaviors, and learning-related outcomes. 


In this paper, we address this gap through a randomized controlled experiment comparing four configurations of the same AI TA. We vary \textit{pedagogical guidance style} by comparing a Socratic AI TA that guides students through questions with a Direct AI TA that provides brief explanations while still withholding complete solutions. We vary \textit{context awareness} by comparing an AI TA with full access to the problem statement and the student's current solution with an AI TA that has access only to the chat history. We investigate students' learning experiences with these AI TAs by answering the following research question:

\begin{enumerate}
\item[\textbf{RQ:}] To what extent does an AI TA's \emph{pedagogical guidance style} and \emph{context awareness} affect introductory programming students' learning experiences, as measured through their (a) \emph{perceptions}, (b) \emph{behaviors}, and (c) \emph{task comprehension}?
\end{enumerate}


We conducted a 2$\times$2 factorial randomized controlled experiment with 132 students in an introductory Python course. Students completed one code-writing and two code-debugging tasks, and each student was randomly assigned to interact with one of four AI TAs configured with a specific combination of pedagogical style and context. We analyzed students' perceptions from a post-survey, interaction behaviors from tool logs, and task comprehension from a post-task explanation of a debugging solution. Notably, the AI TA that combined Socratic guidance with full context was rated least favorably for supporting task completion, showing that pedagogically motivated designs are not automatically experienced by students as more helpful. Our findings provide evidence-based guidance for educators and developers seeking to deploy such tools effectively in real classroom settings.

\section{Related Work}

Recent work in computing education exploring the use of large language models (LLMs) can be broadly grouped into two strands. The first examines the pedagogical capabilities and limitations of LLMs, including their instructional soundness, learning implications, and ability to generate programming tasks or feedback \cite{raihan_2025_llm_csed_review, sarsa_2022_auto_generation, 2023ExpRespofLLMHellas}. The second investigates student reception and adoption of scaffolded, pedagogically grounded LLM systems, including AI TA tools that are deliberately constrained to promote productive learning behaviors \cite{liu2024cs50, liffiton_2024_codehelp, kazemitabaar_2024_codeaid, 2025BotReportZamfirescu, Akcapinar_2024_chatbot, li2025coderunner, 2024CharsAITeachingDenny}. Evaluations of systems such as CodeHelp \cite{liffiton_2024_codehelp} and CodeAid \cite{kazemitabaar_2024_codeaid} are well known examples of this latter strand of work.  There is much interest in this space even beyond computing education, where similar systems have been developed, deployed, and evaluated across disciplines such as chemistry \cite{wang_2025_chemistry_ai}, medicine \cite{fan2025application}, architecture \cite{wang_2025_ai_architecture}, engineering \cite{shi_2026_ee_ta, ter_2026_socratic_chatbot}, and physics \cite{alipour_2025_ai_ta}.
As such systems become more common, there is a need for research that moves towards understanding how students interact with AI TAs, and especially how particular design choices shape those interactions.


One important design dimension is \textit{pedagogical guidance style}. Socratic questioning is pedagogically appealing because it guides students toward discovering solutions themselves rather than providing direct answers \cite{vygotsky2012thought}. Prior work in computing education has shown that Socratic techniques can support learning \cite{tamang_2021_socratic}, and recent studies have explored whether LLMs can generate Socratic guidance for programming tasks \cite{al-hossami_2024_socratic_benchmarking, liding_2026_socratic_hints} and whether Socratic AI TAs can improve learning and retention \cite{tran_2026_pacing}. For instance, Tran et al. found that slower-paced, Socratic-style AI TA interactions produced greater learning gains, particularly among students with less prior knowledge \cite{tran_2026_pacing}. They also found that faster-paced interactions provided initial learning benefits for more advanced students, but these gains were accompanied by reduced retention. Our work complements Tran et al. by extending the study of AI TA guidance beyond interaction pacing and learning/retention outcomes to examine how pedagogical guidance style and context availability shape students' perceptions, interaction behaviors, and immediate post-task comprehension. Additionally, previous work has found that students’ reception of Socratic feedback is mixed \cite{ahmed_2025_ai_ta}. For example, Ahmed et al. found that students perceived Socratic hints as less helpful, less accurate, and more time-consuming than other feedback approaches, even when experts viewed the feedback more favorably \cite{ahmed_2025_ai_ta}. This suggests a tension between pedagogical intent and student experience that warrants further investigation in conversational AI TA settings. 

A second important design dimension is \textit{context awareness}. Instructors have expressed a strong preference for customizing AI chatbots to fit their course-specific needs, including how and where the chatbot is deployed \cite{hou_2026_bespoke_bots}. In programming courses, context may include the problem statement, the student's current code, and the history of the interaction. Providing this context may allow an AI TA to offer more relevant support, while withholding it may encourage students to articulate their problem more explicitly. Prior work has generally reported AI TA systems with fixed configurations (e.g. \cite{diaz2025edugator, liu2024cs50, feng2026bot}) rather than isolating how specific configurations affect student perceptions and learning. Our study addresses this important gap by experimentally comparing AI TAs that vary across both pedagogical guidance style and context awareness.

\section{Methods\protect\footnotemark} 
\footnotetext{Sections of this paper were refined using GenAI tools (ChatGPT and Claude) to format tables \& figures, improve grammar, and clarity.}


\subsection{Study Design}
Our study employed a between-subjects, 2$\times$2 factorial parallel-arm randomized design \cite{mackenzie2024human} to examine the impact of AI TAs' \textit{pedagogical guidance style} (Socratic vs. Direct instruction) and \textit{context awareness} (Full Context of the problem statement and student's current solution vs. No Context) on students' learning experiences. These experiences were operationalized through student perceptions, interaction behaviors, and task comprehension. The decision to manipulate these dimensions was motivated by prior work identifying task context and pedagogical strategy as important configurations preferred by instructors in AI TAs \cite{hou_2026_bespoke_bots}.

Students were asked to solve three programming problems (one code-writing and two code-debugging tasks) on a web-based coding platform with assistance from one of four randomly assigned AI TAs. Following completion of the assignment, all participants completed a survey that gauged their perceptions of the AI TA, their help-seeking behavior, and their comprehension of one of the debugging tasks. The study's methodology and data collection procedures were reviewed and approved by the university's Institutional Review Board (IRB).

    
    

\begin{figure*}[t]
\centering
\includegraphics[width=\textwidth]{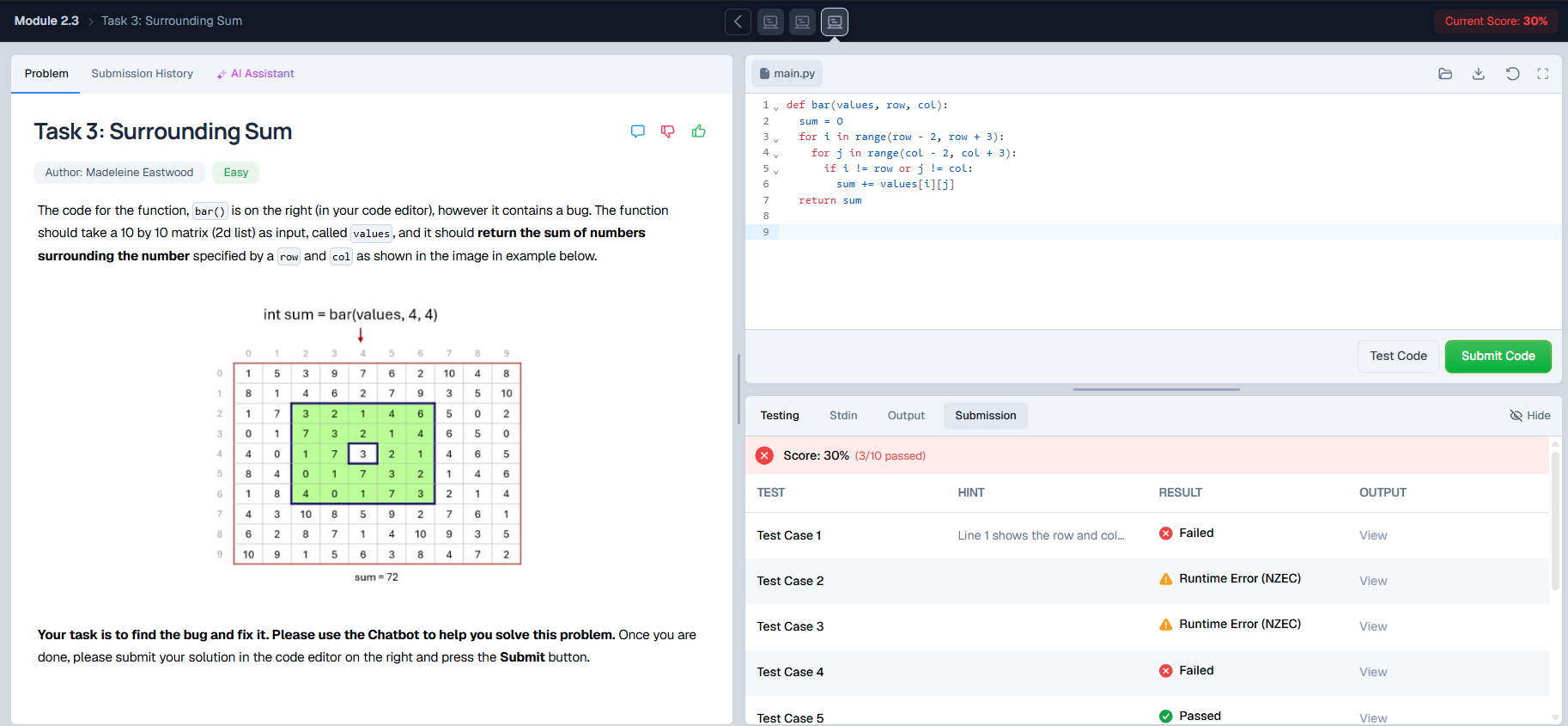}
\caption{Student-facing Interface of the Web-based Coding Platform}
\Description{Screenshot of the coding tool showing the problem statement, submission history, AI assistant, code editor, and test results.}
\label{fig:tool}
\end{figure*}

\subsection{Coding Tool and AI TA design}
\label{tool}

The study was implemented using a coding platform, which allows instructors to author and host programming questions (similar to \cite{2025GuardrailsBehaviorsKapoor}.  A screenshot of the web-based platform is shown in Figure~\ref{fig:tool}.  The application has a split-screen interface with the problem statement, student submission history, ``AI assistant'' on the left, and an embedded text editor on the right. Students could run their own tests or submit their code to run against the instructor-defined test suite.

AI TAs were backed by the OpenAI GPT-4o LLM and returned responses to student queries based on custom prompts with pedagogical guardrails similar to CodeHelp \cite{liffiton_2024_codehelp}. The four AI TAs differed only in their system prompts and contextual information; the interface presented to students was otherwise identical. In the Full Context conditions, the chatbot's system prompt was supplied with the problem statement, the student's current code, and the chat history, whereas in the No Context conditions, it was supplied only with the student's chat history. In the Direct conditions, the chatbot was instructed to provide brief, direct responses without producing complete code solutions, whereas in the Socratic conditions, it was instructed to guide students exclusively through targeted questions without providing direct answers. These two manipulations resulted in four AI TA configurations: (1) Direct + Full Context, (2) Direct + No Context, (3) Socratic + Full Context, and (4) Socratic + No Context.



\subsection{Participants}
Students were recruited from an introductory Python programming course (CS1) at a large public research university in the United States during the Fall 2025 semester. Enrolled students  ($N = 598$) were not familiar with the coding platform used in the study, 
but were invited to participate in our study through in-person and learning management system announcements. They were offered 0.5\% extra credit toward their course grade for completing three programming tasks and a post-survey, and could receive the incentive while opting to exclude their data from research analysis. Data were collected during the final two weeks of the semester, by which point students were familiar with Python lists, loops, and conditionals.


\begin{table}[ht!]
\centering
\caption{Participant Distribution across $2 \times 2$ Experimental Conditions ($N = 132$)}
\label{tab:participant_distribution}
\renewcommand{\arraystretch}{1.0}
\begin{tabular}{@{} l c c c @{}}
\toprule
& \multicolumn{2}{c}{\textbf{Context Awareness}} & \\
\cmidrule(lr){2-3}
\textbf{Feedback Style} & \textbf{Full Context} & \textbf{No Context} & \textbf{Row Total} \\
\midrule
Direct Feedback   & 37 & 32 & 69 \\
Socratic Feedback & 30 & 33 & 63 \\
\midrule
\textbf{Column Total} & 67 & 65 & \textbf{132} \\
\bottomrule
\end{tabular}
\end{table}

Of the 598 students enrolled in the course, 254 completed the intent-to-participate pre-survey. After excluding non-consenting students, incomplete responses, students who did not use the chatbot, and cases with missing or inconsistent chatbot logs, the final analytic dataset consisted of 132 consenting students. Although participants were randomly and evenly assigned to the four conditions at the pre-survey stage, some imbalance occurred because not all students who initially expressed interest completed the study. The final distribution of participants across conditions is shown in Table~\ref{tab:participant_distribution}. Prior programming experience was broadly distributed: 24\% reported being Experienced or Very experienced, 50\% reported Average experience, and 25\% reported being Inexperienced or Very inexperienced, using a self-estimation measure validated in prior work \cite{feigenspan_measuring_pp_2021}.

\begin{table*}[htbp]
\centering
\caption{Research Question Operationalization: Measures, Survey Items/Log Metrics, and Analysis Methods}
\label{tab:metrics}
\scalebox{0.8}{%
\begin{tabular}{@{}p{3.5cm}p{2.8cm}p{10.2cm}p{3cm}@{}}
\toprule
\textbf{Experience Dimension} & \textbf{Measures} & \textbf{Survey Item/Log Metric} & \textbf{Analysis Method} \\
\midrule
(a) Perceptions & Task completion$^{\dag}$ & The [platform] chatbot helped me complete the tasks successfully. & Kruskal-Wallis H test \\
 & Task comprehension$^{\dag}$ & The [platform] chatbot helped me better understand the problem statements. & Kruskal-Wallis H test \\
 & Correct information$^{\dag}$ & The [platform] chatbot provided me with correct information. & Kruskal-Wallis H test \\
 & Stressful interaction$^{\dag}$ & My experience using the [platform] chatbot was stressful. & Kruskal-Wallis H test \\
\addlinespace
\midrule
(b) Behaviors & Used another LLM$^{\#}$ & Did you use another LLM or chatbot? & Chi-square test \\
 & Reason for LLM use$^{\ddag}$ & Why did you use another LLM or chatbot? \textit{(if answered `Yes' to previous question)}& Inductive coding \\
  & Average no. of chats $^{\*}$ & Log data & Kruskal-Wallis H test \\
  & Average chat length $^{\*}$ & Log data & Kruskal-Wallis H test 
  \\
  
\addlinespace
\midrule
(c) Task comprehension & Recall solution$^{\ddag}$ & 
In the tasks you completed, you were asked to debug a function \texttt{bar()} that takes in a 10$\times$10 matrix (2D List) for input, then gives the sum of numbers surrounding a specific cell. Below is the code snippet provided, with the error still present.\par\vspace{4pt}
\texttt{def bar(values, row, col):}\par
\texttt{\phantom{xxxx}sum = 0}\par
\texttt{\phantom{xxxx}for i in range(row-2, row+3):}\par
\texttt{\phantom{xxxxxxxx}for j in range(col-2, col+3):}\par
\texttt{\phantom{xxxxxxxxxxxx}if i != row or j != col:}\par
\texttt{\phantom{xxxxxxxxxxxxxxxx}sum += values[i][j]}\par
\texttt{\phantom{xxxx}return sum}\par
\vspace{4pt}
Please re-identify the error and briefly explain your solution to it.
 & Inductive coding \\
\bottomrule
\end{tabular}%
}
\vspace{4pt}
\par\noindent\small\textit{Note.} Scale used: $^{\dag}$5-point Likert scale (1 = Strongly Disagree, 5 = Strongly Agree). $^{\#}$Binary (Yes/No). $^{\ddag}$Open-ended response.
\end{table*}


\subsection{Study Tasks}
\label{tasks}
We selected three tasks that had been clearly described in prior studies on AI TAs in computing education \cite{2024CharsAITeachingDenny, 2025GuardrailsBehaviorsKapoor}.  The first task was a code-writing task in which students completed a function that determined whether a value was prime. The second task was a debugging task in which students fixed a function intended to determine whether a list contained repeated values. The third task was a debugging task in which students fixed a function that summed values surrounding a specific cell in a 10$\times$10 matrix; the provided implementation had a boundary-checking bug. Function names were anonymized across tasks to reduce the extent to which the chatbot could infer the intended solution from the function name, which was important for evaluating the role of context awareness.

Each student was randomly invited to one of four course shells in our tool, each with a different AI TA but the three identical tasks. Students were given one week to complete the study. Each task included ten instructor-provided test cases, and students were instructed to use the chatbot to help solve the problems.

Most students (128 out of 132) completed all three tasks, passing all test cases. Across all conditions, the average score was 2.98 out of 3 ($median = 3, min = 1, max = 3, \sigma = 0.19$). Students initiated a total of 2,467 messages to the AI TA across the three problems ($\sim$6 messages per student per problem).

\subsection{Data Collection and Analysis}
Data were collected from a post-survey and tool logs that recorded task completion and AI TA interaction behavior. Survey questions were informed by prior work in the community \cite{liffiton_2024_codehelp, 2024CharsAITeachingDenny}. Perception measures used 5-point Likert-scale items assessing whether the chatbot helped students complete the tasks, understand the problem statements, receive correct information, and whether the interaction was stressful. Behavioral measures included whether students used another LLM or chatbot, their reasons for doing so, the number of chat messages sent to the AI TA, and average message length. Task comprehension was assessed through an open-ended post-task response in which students re-identified and explained their solution to the boundary-checking bug from the matrix debugging task.  Table~\ref{tab:metrics} summarizes how each research question dimension was operationalized, including the corresponding survey items or log metrics and analysis methods.

\vspace{3px}
\noindent\textbf{Quantitative Analysis.} Quantitative survey and log data were analyzed using descriptive statistics, followed by inferential tests. Differences across the four experimental conditions were examined using the non-parametric Kruskal-Wallis H test, chosen because the Likert-scale ordinal responses and numerical log data did not follow a normal distribution (Shapiro-Wilk test, $p < .001$). Students' use of another LLM was analyzed using a chi-square test. The significance level was set at $\alpha = .05$ for all tests.

\vspace{3px}
\noindent\textbf{Qualitative Analysis.} Responses to the open-ended questions on task comprehension and students' reasons for using an alternative LLM were analyzed using inductive content analysis \cite{saldana2011fundamentals}. Student responses were inductively coded by two of the authors, with each author coding one of the open-ended questions. The initial codes were iteratively grouped and abstracted into higher-order categories. Two researchers reviewed all codes and categories to verify their consistency, with discrepancies resolved through discussion.

\section{Results}

\subsection{Perception Measures}

\begin{figure*}
    \centering
    \includegraphics[width=0.70\linewidth]{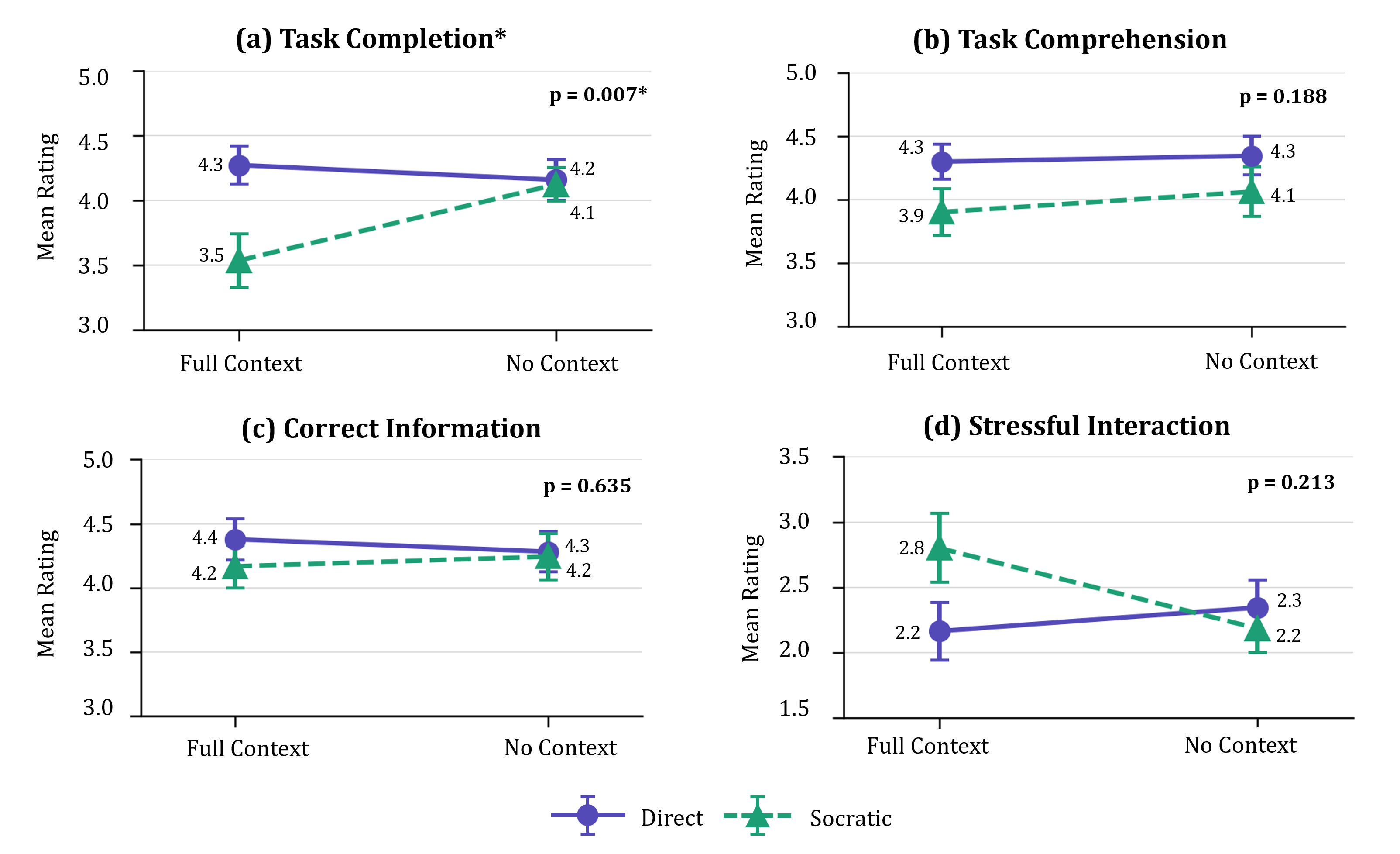}
    \caption{Mean Student Ratings for AI TAs by Interaction of Context Awareness and Pedagogical Guidance Style (N=132)}
    \par\noindent\small\textit{Note.} Ratings measured on a 5-point Likert scale (1 = Strongly Disagree, 5 = Strongly Agree).
    \label{fig:perception_means}
\end{figure*}

A Kruskal--Wallis H test revealed a statistically significant effect of condition on perceived support of the AI TA for \emph{task completion} ($\chi^2(3) = 12.14$, $p = .007$, $N = 132$). Students in the Socratic + Full Context condition rated the AI TA lowest on this measure (mean rank = 48.63, $\mu = 3.53$), substantially below the other three conditions: Direct + Full Context (mean rank = 76.85, $\mu = 4.27$), Direct + No Context (mean rank = 70.75, $\mu = 4.16$), and Socratic + No Context (mean rank = 67.02, $\mu = 4.12$), as shown in Figure~\ref{fig:perception_means}.

Differences across the four conditions were not statistically significant for perceived support for \emph{task comprehension} ($\chi^2(3) = 4.79$, $p = .188$), provision of \emph{correct information} ($\chi^2(3) = 1.71$, $p = .635$), or \emph{stressful interaction} ($\chi^2(3) = 4.50$, $p = .213$). Descriptively, however, students in the Direct conditions rated the AI TA higher on task comprehension than those in the Socratic conditions, and students in the Socratic + Full Context condition rated their interaction as more stressful than students in the other three conditions.

\begin{tcolorbox}[
enhanced, frame hidden,
borderline west={2pt}{0pt}{blue!60!black},
colback=gray!8,
left=10pt, right=8pt, top=6pt, bottom=6pt,
sharp corners
]
\textbf{Perception findings:} Students assigned to a Socratic AI TA with full context rated the AI TA least favorably across the perception measures. They reported significantly lower perceived support for \emph{task completion}, and descriptively the lowest support for \emph{task comprehension} and the highest \emph{stressful interaction} rating.
\end{tcolorbox}

\subsection{Behavior}

\subsubsection{Using another LLM}

Overall, 15\% of students (20 of 132) reported using another LLM or chatbot during the experiment. Although usage rates differed descriptively across conditions, the differences were not statistically significant ($\chi^2(3)=2.47$, $p=.48$). Students in the Socratic + Full Context condition reported the highest rate of external LLM use (23\%, 7 of 30 students), compared with 9--15\% in the other three conditions.

We coded students' open-ended explanations for why they used another LLM. Their reasons were abstracted into three not mutually exclusive categories: \emph{inadequate guidance} ($n=17$), \emph{going in circles} ($n=4$), and \emph{forgetting context} ($n=2$). The most common concern was that the chatbot did not provide enough support, gave explanations that were too general or inflexible, or failed to work dynamically with the student. For example, one student in the Socratic + Full Context condition wrote that the chatbot \textit{``doesn't really expand or work with the student very well''} and was ``\textit{super rigid}''. Another student in the Socratic + No Context condition described the chatbot as using ``\textit{circle reasoning}'', where it repeatedly answered questions with more questions rather than helping with the code.

\subsubsection{Interaction volume}

A Kruskal--Wallis H test indicated a statistically significant difference in the number of queries students sent to the AI TA across the four conditions ($\chi^2(3) = 13.28$, $p = .004$). Students in the Socratic + No Context condition sent substantially more queries ($\mu = 11.1$ per problem) than students in the Socratic + Full Context ($\mu = 5.1$), Direct + Full Context ($\mu = 4.1$), and Direct + No Context ($\mu = 4.7$) conditions. This suggests that students who received Socratic guidance without problem context required more back-and-forth with the chatbot to make progress.

A Kruskal--Wallis H test also revealed a significant difference in average message length across conditions ($\chi^2(3) = 29.96$, $p < .001$). Students in the Direct + No Context condition wrote the longest messages on average ($\mu = 126$ characters per message), followed by Direct + Full Context ($\mu = 81$), Socratic + No Context ($\mu = 69$), and Socratic + Full Context ($\mu = 54$). This pattern suggests that when the chatbot lacked context, students often had to describe their situation in more detail before receiving useful support.

\begin{tcolorbox}[
enhanced, frame hidden,
borderline west={2pt}{0pt}{blue!60!black},
colback=gray!8,
left=10pt, right=8pt, top=6pt, bottom=6pt,
sharp corners
]
\textbf{Behavior findings:} Behavioral engagement differed significantly across conditions in query volume and message length. Students in the Socratic + No Context condition sent the most queries per problem, while students in the Socratic + Full Context condition sent the shortest messages on average. Descriptively, students in the Socratic + Full Context condition also reported the highest, though non-significant, rate of external LLM use.
\end{tcolorbox}

\begin{figure}[ht]
\centering
\includegraphics[width=0.79\columnwidth]{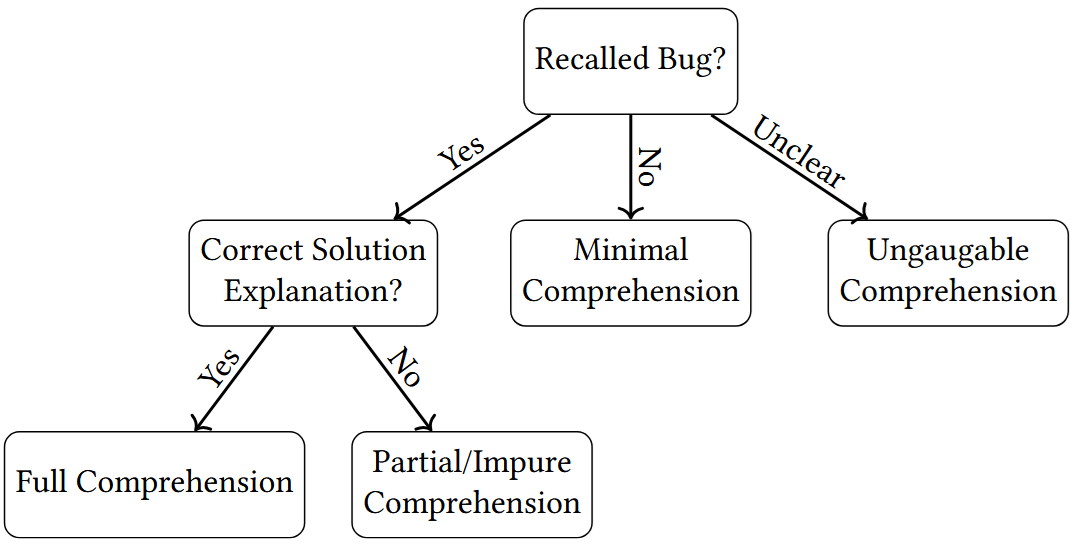}
\caption{Decision Flowchart for Task Comprehension}
\Description{Decision flowchart for task comprehension categories}
\label{fig:flowchart}
\end{figure}

\subsection{Task Comprehension Analysis}
We analyzed students' responses to a post-task survey question, administered immediately after task completion, asking them to re-identify the error and explain their solution for Task 3 (``Surrounding Sum''). This assessment reflects students' immediate comprehension of the problem alongside their own debugging solution, rather than a longer-term retention of the underlying concept, and allowed us to examine how the AI TA's guidance style and context awareness may have shaped that comprehension. Responses were coded into four comprehension categories (see Figure \ref{fig:flowchart}). \emph{Full comprehension} required students to identify the indices out-of-bounds bug and explain a correct fix. \emph{Partial/impure comprehension} captured responses that identified the correct bug but had incomplete, unclear, or faulty reasoning. \emph{Minimal comprehension} captured responses that did not identify the out-of-bounds bug. \emph{Ungaugable comprehension} captured responses where it was unclear whether they had identified the bug.  


\begin{figure}[t]
\centering
\includegraphics[width=1.0\columnwidth]{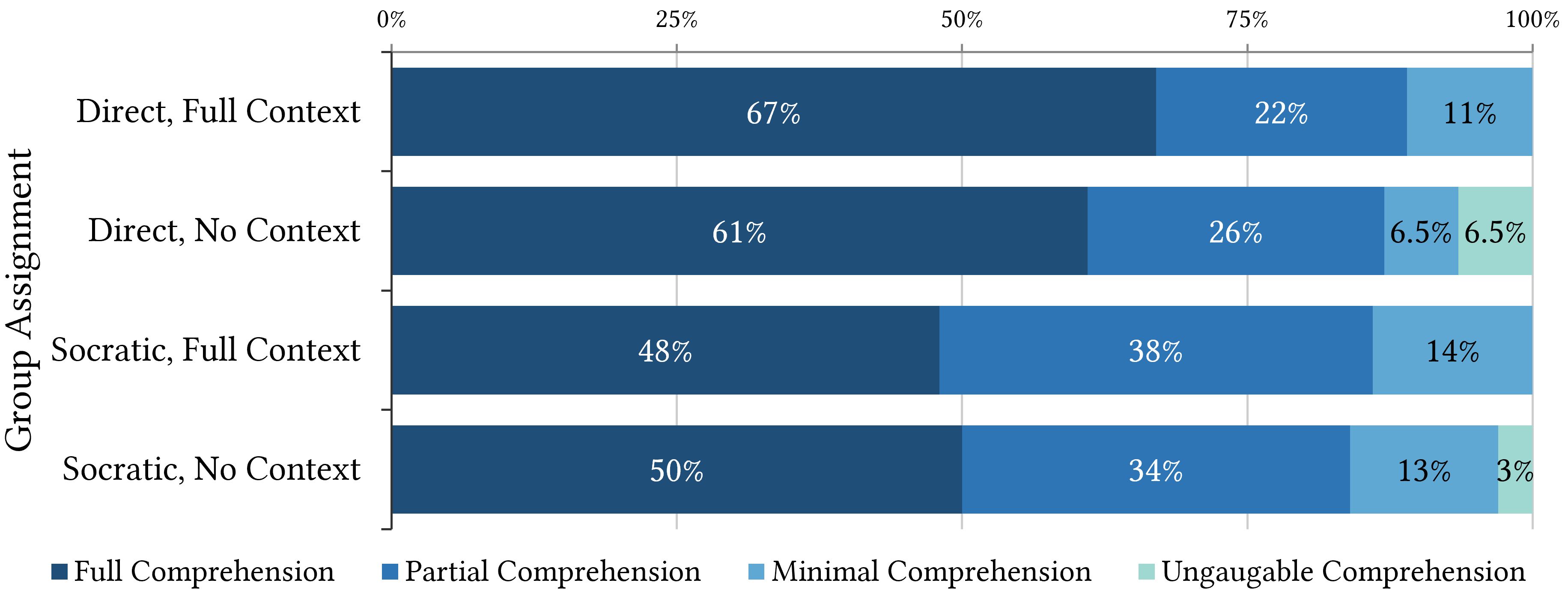}
\caption{Comprehension Level by Group Assignment}
\Description{Comprehension Level by Group Assignment}
\label{fig:comprehension}
\end{figure}

Across all non-blank responses ($n = 128$), 57\% students showed full comprehension, 30\% showed partial/impure comprehension, 11\% showed minimal comprehension, and 2\% were ungaugable. When broken down by condition, the Direct + Full Context group had the highest proportion of full-comprehension responses (67\%), followed by Direct + No Context (61\%), Socratic + No Context (50\%), and Socratic + Full Context (48\%). The Socratic + Full Context group also had the highest proportion of minimal-comprehension responses (14\%, see Figure \ref{fig:comprehension}). Overall, comprehension patterns appeared to differ more by pedagogical guidance style than by context awareness: Direct conditions produced a higher average proportion of full-comprehension responses than Socratic conditions.

\begin{tcolorbox}[
enhanced, frame hidden,
borderline west={2pt}{0pt}{blue!60!black},
colback=gray!8,
left=10pt, right=8pt, top=4pt, bottom=4pt,
sharp corners
]
\textbf{Task comprehension findings:} Descriptively, students in the Direct conditions showed a higher proportion of responses that demonstrated full-comprehension than students in the Socratic conditions. The Direct + Full Context condition had the highest proportion of full-comprehension responses, while the Socratic + Full Context condition had the lowest.
\end{tcolorbox}

\section{Discussion}
\noindent
\textbf{Role of Pedagogical Guidance Style}: Our results suggest that the AI TA's guidance style shaped how students engaged with the tool. Students in the Socratic conditions sent the most queries, suggesting that a chatbot that responds with guiding questions rather than direct answers can require more back-and-forth before students make progress. In this sense, Socratic guidance appeared to increase interaction volume as observed in other domains outside of computing \cite{ter_2026_socratic_chatbot}, especially when the chatbot lacked task context.  


However, the perception and comprehension results tell a more cautionary story. Students rated the Socratic + Full Context condition least favorably for supporting task completion, and this condition also had the lowest proportion of full-comprehension post-task explanations. This raises an important tension for educators considering AI TA deployment.  A pedagogically motivated approach that encourages students to reason through problems independently may still be experienced as frustrating or unhelpful (similar to results from Ahmed et al. \cite{ahmed_2025_ai_ta}), especially when students are focused on completing a programming task. Educators should therefore consider whether the goal of a given assignment is task completion, conceptual understanding, or productive struggle, and configure the AI TA's guidance style accordingly.

\vspace{5px}
\noindent
\textbf{Role of Contextual Awareness}: Context awareness also played a notable role in shaping student behavior. Students in the No Context conditions, where the chatbot had access only to the chat history and not to the problem statement or student's current code, tended to write longer individual queries than those in the Full Context conditions. This suggests that when the chatbot lacked information about the problem, students compensated by providing more detail in their messages, essentially doing the work of explaining their situation before the chatbot could help them.


At the same time, Full Context did not guarantee a better student experience. The Socratic + Full Context condition produced the shortest messages on average, suggesting that students did not need to explain much before receiving a response. Yet this same condition was rated least favorably for task completion and had the lowest proportion of full-comprehension responses on the post-task. This points to a practical implication for educators, which is that providing an AI TA with full task context may reduce the effort required to formulate queries, but contextual awareness alone does not ensure that the support will feel useful or lead to stronger understanding. Instructors deploying AI TAs in programming courses should therefore consider integrating the tool with the assignment environment while also carefully designing how the chatbot uses that context pedagogically.

\vspace{5px}
\noindent
\textbf{Interaction Between Pedagogical Guidance Style and Context Awareness}: The interaction plots in Figure~\ref{fig:perception_means} show why guidance style and context awareness should not be interpreted independently.  When Direct guidance was available, students rated the chatbot similarly for supporting task completion regardless of how much context was provided.  In contrast, when the chatbot was Socratic, providing full context reduced students' perceived support for task completion. This suggests that context awareness was not uniformly beneficial, and that its effect depended on the pedagogical style with which it was paired.

The behavioral results reinforce this interpretation. The Socratic + No Context condition produced the highest query volume, which makes intuitive sense because when a chatbot avoids direct answers and also lacks knowledge of the task, students must work harder to get useful help. The result may be a high-effort interaction that feels more exhausting than educational, particularly for students who are already struggling. In contrast, the Direct + Full Context condition produced the lowest query volume, suggesting that students could get help quickly when the chatbot provided direct guidance and already had access to the problem and their current code.

These findings suggest that guardrails can become roadblocks when pedagogical guidance and contextual awareness are not well balanced. 
A Socratic chatbot with limited context may demand too much effort from students, while a highly contextualized chatbot that remains rigidly Socratic may still fail to provide the kind of support students find useful, as reflected in the weaker descriptive post-task comprehension pattern for the Socratic + Full Context condition. Rather than treating ``more scaffolding'' or ``more context'' as inherently better, educators and system designers should consider AI TAs as configurable learning supports whose effectiveness depends on the fit between the task, the student's needs, and the form of help provided. Future work should explore adaptive AI TAs that can shift between Socratic and more direct guidance based on students' progress, confusion, or help-seeking behavior.

\vspace{5px}
\noindent
\textbf{Limitations}: Our study was conducted in an introductory programming course at a large research university. Although this controlled setting supports high internal validity, it may limit the generalizability of the findings to other contexts. Another limitation is that we measured task comprehension for only one task, a design choice we made to reduce survey fatigue. This design captures immediate comprehension rather than knowledge retention.  Future studies could investigate the impact of AI TA configurations on long-term knowledge retention. Finally, qualitative coding of open-ended responses involves some degree of subjectivity; we mitigate this by developing a codebook, transparent reporting of our process, and by providing a detailed flowchart of our approach.

\section{Conclusion}

AI TAs are increasingly being deployed in programming courses to provide scalable help while using pedagogical guardrails to avoid simply giving students complete solutions. However, our findings show that not all guardrailed designs are experienced by students as helpful --- their effect depends on \emph{pedagogical guidance style} and \emph{context awareness}. In this study, we moved beyond evaluating a single AI TA system as a complete intervention (similar to \cite{kazemitabaar_2024_codeaid, li2025coderunner,liffiton_2024_codehelp, liu2024cs50}) by experimentally comparing four configurations of the same AI TA, varying both guidance style and context awareness. Across students’ perceptions, interaction behaviors, and post-task comprehension, the Socratic + Full Context condition showed the least favorable descriptive pattern. Students rated it significantly lower in perceived support for task completion, and it also showed less favorable descriptive patterns for perceived comprehension support, higher interaction stress, greater external LLM use, and lower post-task comprehension. These results highlight a key design challenge for AI TAs in programming courses, which is that pedagogically motivated configurations can become roadblocks if students experience them as too rigid or insufficiently useful. Effective AI TA design will therefore require balancing pedagogical guidance with contextual awareness in ways that support learning while keeping students engaged with course-approved tools.



\balance

\bibliographystyle{ACM-Reference-Format}
\bibliography{9_references}

\end{document}